\documentclass[12pt]{article}

\usepackage[usenames]{color}
\usepackage{ifthen}
\usepackage[authoryear]{natbib}
\usepackage{graphicx}
\usepackage{longtable}
\usepackage{booktabs}
\usepackage{latexsym}
\usepackage{amssymb}
\usepackage{amsmath}
\usepackage{amsfonts,amsthm,enumerate,array}
\usepackage{bm,multicol}
\usepackage{caption}
\usepackage[hidelinks,colorlinks=true,linkcolor=cyan,
            citecolor=cyan, urlcolor=cyan]{hyperref}
\usepackage{chngcntr} 
\usepackage{mathrsfs} 
\usepackage{adjustbox}
\usepackage{multirow}
\usepackage{enumitem}
\usepackage{algorithm2e}
\usepackage{bbm}
\usepackage{setspace}
\usepackage{rotating}
\usepackage{mathabx}
\usepackage{ulem}
\usepackage{authblk}

\makeatletter 
\newcommand{\labitem}[2]{%
\def\@itemlabel{\textbf{#1.}}
\item
\def\@currentlabel{#1}\label{#2}}
\makeatother

\makeatletter 
\newcommand{\labitemc}[2]{%
\def\@itemlabel{\textbf{#1}}
\item
\def\@currentlabel{#1}\label{#2}}
\makeatother

\newboolean{tab}
\setboolean{tab}{false}

\makeatletter\@addtoreset{equation}{section}\makeatother

\renewcommand{\tilde}{\widetilde}
\renewcommand{\hat}{\widehat}

\DeclareMathOperator{\diag}{diag}

\advance\topmargin by -0.85truein
\advance\textheight by 1.25truein

\tiny\normalsize
\date{}
\begin{document}
\thispagestyle{empty}
  \title{Bayesian Unit-level Quantile Modeling with Application to Wealth Data}
  \author[1]{Daniel Vedensky\thanks{Corresponding author, \href{mailto:daniel_vedensky@sfu.ca}{\texttt{daniel\_vedensky@sfu.ca}}}}
  \author[2]{Paul A. Parker}
  \author[3]{Scott H. Holan}
  \affil[1]{Simon Fraser University}
  \affil[2]{University of California, Santa Cruz}
  \affil[3]{University of Missouri}
 \maketitle 

 \begin{abstract}
Modeling wealth quantiles over time and space and across fine-grained
demographic classifications is of vital importance to policy makers.
However, such data present unique challenges for standard small area
estimation methodologies because they are highly skewed and heavy-tailed with
both positive and negative values.
Traditional area-level modeling and design-based approaches struggle to
capture these features.
Meanwhile, existing unit-level methodology frequently treats the survey
design as ignorable and does not apply to surveys that are repeated over time.
Often the analysis is restricted to alternative response variables that are
non-negative, such as poverty indicators or income.
To overcome these limitations, we propose a Bayesian, asymmetric Laplace
pseudolikelihood model that is able to capture the major features of wealth
while adjusting for survey weights and accounting for both spatial and
temporal dependence.
We present a Gibbs sampling algorithm for our model, as well as a variational
Bayes algorithm for fast computation.
We apply the method to the Survey of Income and Program Participation to characterize geographic, temporal, and
demographic variation in the distribution of U.S. household wealth, including for sparsely
sampled domains where direct estimates are unavailable or unstable.
 \end{abstract}
 

\newcommand{\areamedianfigureheight}{0.84\textheight}
\section{Introduction}\label{sec:intro} 
Measuring economic variables at an individual or household level rather than in aggregate is crucial for assessing the economic well-being of a population \citep{stiglitz_2009, oecd_2013, european_central_bank_2019}. 
Household wealth, defined as the sum of household members' assets minus their liabilities, is particularly important for understanding economic health \citep{census2008}.
By analyzing at the household level, it is possible to study the distribution of wealth and therefore assess phenomena like wealth inequality, which has been deemed ``one of the defining challenges of this century.''\footnote{\url{https://www.cfr.org/backgrounder/us-inequality-debate}}
For policy makers, it is especially important to consider the change in wealth distribution over time as this provides insight into the impact of welfare programs and fiscal policies.
Moreover, because household surveys contain detailed demographic data, they allow policy makers to identify the subgroups most in need of assistance from these programs and policies \citep{oecd_2013}.
Recent research therefore emphasizes examining differences in household wealth across age, sex, race, and geographic location when such data are available \citep{balestra_etal_2025}.

As such, there is need to develop methods for household surveys that measure wealth at a fine-grained level.
In the United States, the Survey of Income and Program Participation (SIPP) is one of the primary sources for detailed household-level wealth data.
Data collection is carried out annually, and data are also recorded at the state-level, so comparisons can be made by region.
We describe the structure of SIPP in more detail in the following section and use its structure to motivate our proposed methodology and data analysis. 
However, the challenges in modeling SIPP carry over to any household survey measuring wealth:
there is spatial dependence, temporal dependence, and a complex survey sampling design.
In order to properly model the data, all of these dependencies must be addressed.

In general, wealth is challenging to model due to its heavy-tailed and skewed distribution (e.g., in the U.S., it was recently estimated the top $0.1\%$ of households hold around $20\%$ of all wealth \citep{Saez2020}).
Because wealth takes on extreme values that can be either positive or negative, common remedies like a log transformation do not apply.
As a result, existing statistical analyses tend to perform ad hoc transformations (e.g., truncating the data or dropping negative values) 
or avoid analyzing wealth directly.
It is common to instead examine measures such as wealth (or poverty) indicators that take the form of proportions \citep{Lehtonen2016}, model the data ranks \citep{Chenevert2017}, or examine proxies which do not contain negative values like income or TIP curves \citep{FourrierNicolai2019}.
Notably, such proxies do not capture the same aspects of economic well-being that wealth does: households with high wealth are not necessarily the same as those with high incomes \citep{Bricker2020}.

When wealth is modeled directly, it tends to be through the use of direct estimators, whose estimates are unreliable for domains with small sample sizes. 
Instead, such an analysis typically examines coarse categories, like race \citep{Aladangady2021} or sex \citep{Ruel2012}, but rarely variation over time or across geographies.
In particular, methodology for the temporal aspect is often overlooked, and estimates for each year are computed independently.
Not only does this ignore the dependence across time, which could inflate standard errors, %
but it also precludes analysis of distributional changes over time. 
This is key to answering questions such as whether households are entering and exiting poverty as opposed to remaining there chronically, whether wealth inequality is increasing or holding steady, and so on.

Given the heavy tails and skew of wealth data, it is natural to examine quantiles rather than means.
Quantile regression methods \citep{Koenker1978} have recently been extended to spatio-temporal problems in a number of ways, which would be useful for wealth data.
In the Bayesian literature this generally relies on an Asymmetric Laplace (AL) likelihood \citep{Yu2001} or the more recently developed extended Asymmetric Laplace (exAL) \citep{Yan2025}.
For instance, \citet{Lum2012} introduce an AL stochastic process for spatial data, while \citet{waldmann_2015} propose a variational Bayes (VB) algorithm for geoadditive AL quantile regression.
Their algorithm extends the i.i.d. AL variational approximation of \citet{wand_2011} to geoadditive quantile regression.
\citet{Barata2022} present a dynamic state-space model with exAL errors for time series data that can be sampled with MCMC or VB.
However, these methods are not directly applicable to wealth data obtained under a complex survey design.

Unit-level SAE models provide a natural way to extend spatiotemporal methods to survey data as they allow us to produce estimates for demographic categories across space and time, while also considering person-level covariates and variation across individuals.
Several works have explored this connection, but assume an ignorable sampling design and cross-sectional data.
\citet{molina_2010} propose an empirical best prediction approach for nonlinear finite-population quantities, including quantiles and poverty indicators, under a cross-sectional Gaussian nested-error model with independent area effects.
In a frequentist context, \citet{berg_2019} propose a mixed-effects quantile regression method based on a linearly interpolated generalized Pareto distribution.
They estimate many quantiles jointly over a large grid and aggregate the resulting unit-level distributions to predict finite-population quantiles.
Similarly, \citet{weidenhammer_2016} use an AL working likelihood and estimate a grid of quantiles.
They also simulate from these constructed distributions and aggregate the simulated outcomes to estimate finite-population quantiles.
Whereas \citet{weidenhammer_2016} reconstruct a unit-level distribution from conditional quantile models fitted over a grid, our approach targets selected quantiles within a Bayesian temporal model and incorporates survey weights directly through the pseudolikelihood.
Both the approach of \citet{berg_2019} and that of \citet{weidenhammer_2016} are cross-sectional, assume independent area effects, and do not explicitly account for informative sampling.
An exception to the cross-sectional restriction is the model proposed by \citet{Neelon2015}, who combine an asymmetric Laplace working likelihood with spatial, temporal, and space--time interaction effects. 
However, their method estimates area-specific conditional quantiles rather than finite-population area quantiles and does not account for a complex survey design.

Many techniques exist for accounting for a complex survey design in unit-level models \citep{Parker_2019},
but existing methodology is generally not adequate for modeling detailed wealth data.
As we show in a simulation study, the longitudinal, Gaussian unit-level model of \citet{Vedensky2025} is unable to produce workable estimates due to the heavy tails in the data.
Recently, \citet{Nascimento2024} proposed unit-level methodology for quantile regression of survey data with an AL likelihood, but the method applies only for cross-sectional data and not for performing the task of small area estimation.
Thus, neither the spatiotemporal quantile-regression literature nor the survey unit-level literature provides a method for temporal small area estimation of finite-population wealth quantiles under a complex survey design.
To address this gap, we propose a Bayesian unit-level quantile model that incorporates spatial and temporal dependence, accounts for the complex survey design through a Bayesian pseudolikelihood, and uses posterior predictive simulation and poststratification to estimate finite-population wealth quantiles for demographic groups across areas and over time, with scalable MCMC and VB inference.
To our knowledge, no previous analysis has produced model-based finite-population wealth quantiles for demographic domains across U.S. states and years while accounting for the survey design and borrowing information over space and time.

The rest of the paper proceeds as follows.
Section~\ref{sec:dataset} describes the SIPP dataset.
Section~\ref{sec:methods} describes our proposed methodology.
Section~\ref{sec:sim} presents the results of an empirical simulation that demonstrate the efficacy of the proposed method and
Section~\ref{sec:data_analysis} contains a data analysis of the SIPP data.
We conclude with a discussion in Section~\ref{sec:conclusion}.

\section{Data Description}\label{sec:dataset}
\begin{figure}[ht!]
    \centering
    \includegraphics[width=.8\linewidth]{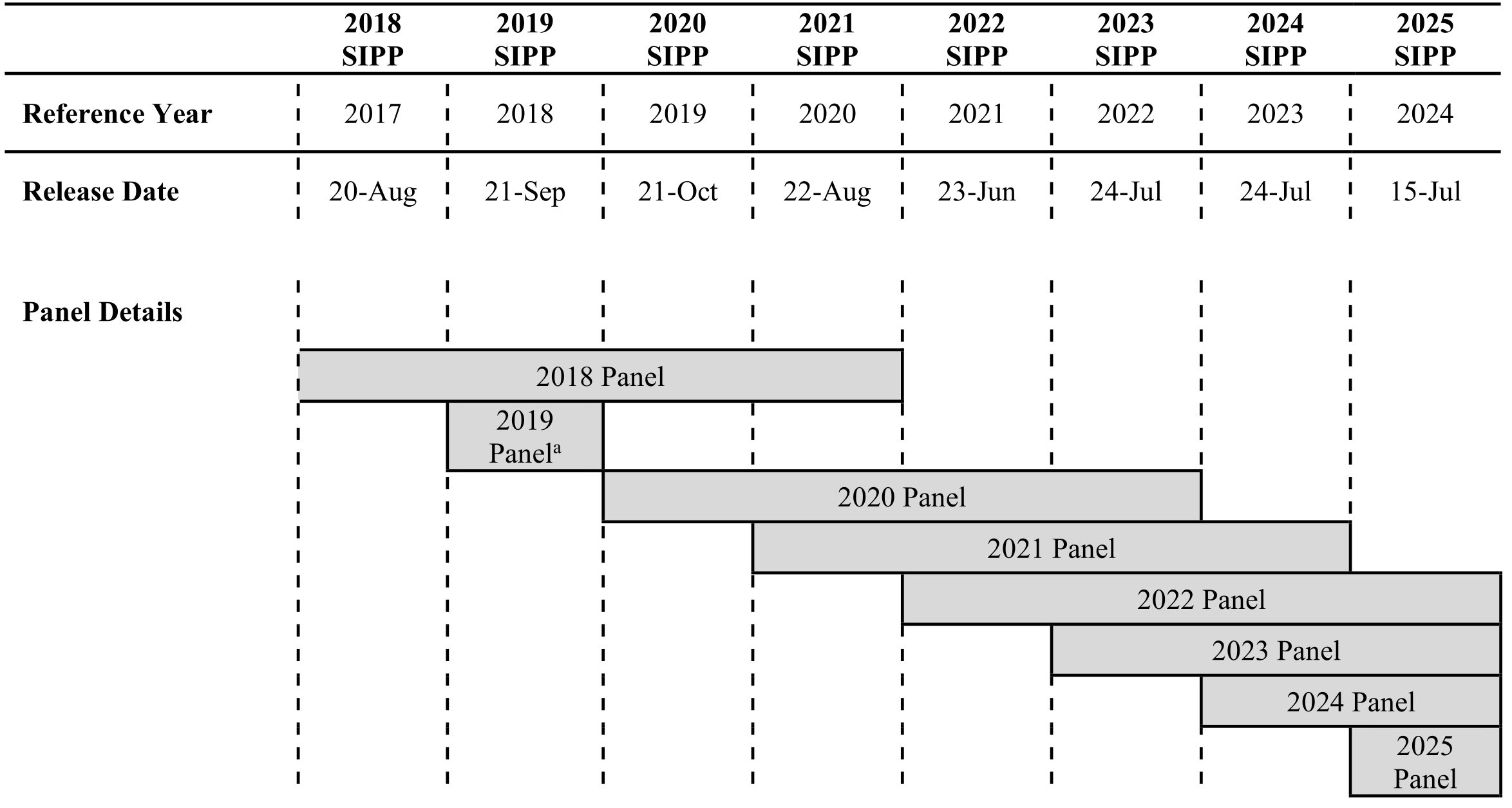}
        \caption{Illustration of SIPP rotating panel design reproduced from the 2025 SIPP User's Guide \citep{sipp_users_guide_2025}. 
    \label{fig:sipp_structure}}
\end{figure}

SIPP is a longitudinal survey that was first fielded in 1984 and continues to be conducted to the present day.
However, a significant redesign was carried out in 2014 \citep{sipp_redesign2018}.
The 2018 SIPP marked the first return to a truly longitudinal design with multiple concurrent panels.
We focus on the subset of the public-use data that has been released since then, with the most recent release the 2025 SIPP.\footnote{Available at \url{https://www.census.gov/programs-surveys/sipp/data/datasets.html}}
Figure~\ref{fig:sipp_structure} summarizes the panel structure the survey has adhered to since 2018.
Interviews for the SIPP ask questions about the preceding year so that the 2018--2025 SIPP releases contain data for 2017--2024.
The longitudinal structure is such that a household is interviewed once a year for four years, with information recorded at a monthly resolution.
Each interview year is referred to as a wave.
Note that data collection for the 2019 panel was disrupted by a lack of funding and the COVID-19 pandemic and the panel was discontinued\footnote{\url{https://www.census.gov/programs-surveys/sipp/tech-documentation/user-notes/2019-usernotes/2019-discont-2019-pnl.html}} \citep{Tolliver2023}.


 SIPP stands out among longitudinal surveys both for its large sample size (373,871 people and 4,430,670 person-months across the period we consider)
and for the range of data it collects \citep{sipp_disclosure2024}. 
The original intent of the survey was to track poverty status, the effect of government programs, and economic well-being, \citep{sipp_redesign2018} so the questionnaire contains detailed economic questions about assets, liabilities, government welfare receipts, and income.

We specifically examine household net worth as a standard representation of wealth.
This is taken to be the sum of all household assets minus all household liabilities.
Note that, if necessary, the SIPP data allow for far more fine-grained analyses, such as net worth excluding home equity, or examining debt alone.

Because net worth is a household-level variable, we group the data by household, taking the demographics of the householder as covariate values.
We restrict our attention to the contiguous United States, which leaves 83,159 unique households and 155,818 household-years.
These data are highly variable with a range of -62,730,996 to 435,045,300 and standard deviation 2,725,710, as well as highly skewed---the median is 165,137.5 while the mean is 678,304.

\section{Methodology}\label{sec:methods}
Let $\mathcal{U}_t = \{1,\ldots, N_t\}$ enumerate a finite population at a given time $t$ that is the union of disjoint subpopulations $\mathcal{U}_{t1},\ldots,\mathcal{U}_{tK}$ such that $\mathcal{U}_t = \bigcup_{k=1}^K\mathcal{U}_{tk}$ and $|\mathcal{U}_{tk}| = N_{tk}$. 
 Each $\mathcal{U}_{tk}$ may represent the population within an area or some demographic classification.
At each time $t=1,\ldots,T$ a sample $\mathcal{S}_{kt} \subseteq \mathcal{U}_{kt}$ of size $n_{kt}$ is taken with known probability $\pi_{ikt}= P(i\in \mathcal{S}_{kt})$ and inverse-probability weights are constructed as $w_{ikt} = 1/\pi_{ikt}$.
Associated with each unit $i$ in area $k$ at time $t$ is a response value $y_{ikt}$ and a $q$-dimensional vector of covariates $\bm x_{ikt}$.

For expository purposes, we will first focus on the cross-sectional case and suppress the $k$ and $t$ subscripts.
Typically, to estimate quantiles of a complex sample, the empirical distribution function \citep{Woodruff1952, Dorfman1993} is first estimated as
\begin{equation}
\hat{F}(u) = \sum_{j=1}^n \frac{w_j}{\sum_{i=1}^n w_i}\mathbf{1}(y_j \leq u)  \label{eq:ecdf}
\end{equation}
and the $p$th quantile is estimated as
\begin{equation}
\hat{q}_p=\min\{y: \hat{F}(y)\geq p\}.
\label{eq:survey_quantile}
\end{equation}
Confidence intervals can then be constructed based on the asymptotic normality of (\ref{eq:ecdf}).
While direct estimators are appealing for their simplicity and properties like design-consistency and asymptotic normality, they cannot produce adequate estimates and confidence intervals for small domains with low sample sizes.
Moreover, unlike Horvitz--Thompson-type estimates, (\ref{eq:survey_quantile}) is not design unbiased \citep{schulzewaltrup_2016}.

In such cases, it is necessary to model the data and borrow strength from similar categories.
A natural choice is quantile regression \citep{Koenker1978} which, in the frequentist setting, proceeds by minimizing
  \[\sum_{i=1}^n\rho_p(y_i-\bm x'_i\bm\beta(p)), \]
where $\rho_p(u) = u(p - I(u<0))$
is the check-loss function.
In the Bayesian setting, quantile regression is commonly implemented using an Asymmetric Laplace (AL) working likelihood
  \begin{equation}
    AL_p(y_i|\mu_i,\sigma) = \frac{p(1-p)}{\sigma}\exp\left\{-\frac{\rho_p(y_i-\mu_i)}{\sigma}\right\},\label{eq:AL_likelihood}
  \end{equation}
  assuming 
  \begin{equation}
      \mu_i=\bm x_i'\bm\beta(p) = q_p(y_i|\bm x_i), \label{eq:conditional_quantile}
  \end{equation} 
  specifying priors for $\bm\beta(p)$ and $\sigma$, and estimating $\bm\beta(p)$ \citep{Yu2001}.
  For median regression ($p=0.5$) the above is equivalent to a Laplace likelihood.

In order to facilitate posterior computation, \citet{Kozumi2011} propose a data augmentation algorithm.
Defining  $\theta=\frac{1-2p}{p(1-p)}$ and $\tau^2 = \frac{2}{p(1-p)}$ they show an $AL_p(\mu, \sigma)$ random variable, $y_i$, can be written
\[y_i = \mu_i + \theta \nu_i + \tau \sqrt{\sigma \nu_i}u_i,\]
where $\nu_i\sim \text{Exp}(\sigma)$ and $u_i\sim \text{N}(0,1)$, 
with Exp$(\sigma)$ parameterized to have mean $\sigma$ and with $\nu_i$ and $u_i$ mutually independent.
That is, a random variable $y_i\sim AL_p(\mu_i, \sigma)$ can be sampled in two steps as 
\begin{align*}
  y_i \mid \nu_i, \sigma &\sim \text{N}(\mu + \theta \nu_i, \tau^2\nu_i\sigma)\\
  \nu_i \mid \sigma &\sim \text{Exp}(\sigma).
\end{align*}

A common way to incorporate survey weights into such a Bayesian model is to replace the likelihood with an exponentially weighted pseudolikelihood \citep{binder_1983, savitsky_toth_2016}.
In the Bayesian case \citep{savitsky_toth_2016}, the survey weights are scaled to sum to the sample size
\[\widetilde{w}_i = n\frac{w_i}{\sum_{i=1}^n w_i}.\]
  For the AL, the resulting pseudolikelihood can be written, up to a factor that does not depend on the model parameters, as \citep{Nascimento2024}
  \[
  \prod_{i=1}^n
  \frac{\widetilde w_i p(1-p)}{\sigma}
  \exp\left\{-\frac{\widetilde w_i\rho_p(y_i-\bm x_i'\bm\beta(p))}{\sigma}\right\}.
  \]
  This working likelihood can be sampled from in the same fashion as before
  \begin{align*}
    y_i \mid \nu_i, \sigma,\bm\beta &\sim \text{N}(\bm x_i'\bm\beta(p) + \theta \nu_i, \tau^2\nu_i\sigma/\widetilde w_i)\\
    \nu_i \mid \sigma &\sim \text{Exp}(\sigma/\widetilde w_i).
  \end{align*}
  Assuming a Gaussian prior on $\bm\beta(p)$, the $q$-dimensional vector of fixed effects, retains conjugacy.

Our proposed extension is threefold.
First, for a cross-sectional model, we add an $r$-dimensional vector of areal random effects, $\bm\eta$, to the conditional quantile function so that the data model is instead $y_i\mid \cdot \sim AL_p(\bm x_i'\bm\beta + \bm\psi_i'\bm\eta, \sigma/\widetilde w_i)$.
To incorporate additional spatial dependence, we take $\bm\psi_i'$ to be an $r$-dimensional vector of basis functions, truncated for dimension-reduction.
Following \citet{parker_holan_janicki_2022}, we use the first 20 eigenvectors of the state adjacency matrix, corresponding to its positive eigenvalues, as our spatial basis functions.
  The full cross-sectional model can then be specified as 
  \begin{align}
    \nonumber
     y_i \mid \cdot &\overset{ind.}{\sim} \text{AL}_p\left(\mu_{i}(p), \sigma/\widetilde w_i\right )\\\nonumber 
  \mu_{i}(p) &= \bm x_i'\bm\beta(p) + \bm\psi_i'\bm\eta(p)\\
  \bm\eta(p) \mid \sigma_\eta^2 &\sim \text{N}(\bm 0, \sigma^2_{\eta} I_r)  \label{eq:AL_cs}\\ \nonumber
  \bm\beta(p) &\sim \text{N}(\bm 0, \sigma^2_\beta I_q)\\\nonumber
  \sigma &\sim \text{IG}(a_\sigma, b_\sigma)\\\nonumber
  \sigma^2_\eta &\sim \text{IG}(a_\eta, b_\eta)
 \end{align}
Here $a_\sigma, b_\sigma, a_\eta, b_\eta>0$ are fixed hyperparameters.
We use the inverse-gamma parameterization with density $f(z\mid a,b)=b^a\{\Gamma(a)\}^{-1}z^{-(a+1)}\exp(-b/z)$ for $z>0$.

Second, to borrow information across time, we allow both the fixed-effect coefficients and spatial basis coefficients to vary according to separate AR(1) processes.
The proposed temporal model is
\begin{align} y_{it}\mid\cdot &\overset{\mathrm{ind.}}{\sim} \mathrm{AL}_p\left( \mu_{it}(p), \sigma_t/\widetilde w_{it} \right), && i=1,\ldots,n_t,\quad t=1,\ldots,T, \nonumber\\
\mu_{it}(p) &= \bm{x}_{it}^{\prime}\bm{\beta}_t(p) + \bm{\psi}_i^{\prime}\bm{\eta}_t(p), \nonumber\\
\bm{\beta}_t(p)\mid\bm{\beta}_{t-1}(p) &\sim \mathrm{N}_q\left( \phi_\beta\bm{\beta}_{t-1}(p), \sigma_\beta^2 I_q \right), && t=2,\ldots,T, \label{eq:AL_lon} \\
\bm{\eta}_t(p)\mid\bm{\eta}_{t-1}(p) &\sim \mathrm{N}_r\left( \phi_\eta\bm{\eta}_{t-1}(p), \sigma_\eta^2 I_r \right), && t=2,\ldots,T. \nonumber
 \end{align}
The survey weights are now scaled separately at each time point to sum to $n_t$.
The initial states and hyperpriors are specified as 
\begin{align} 
\bm{\beta}_1(p) &\sim \mathrm{N}_q\left( \bm{0}, \sigma_{\beta_1}^2 I_q \right), & \bm{\eta}_1(p)\mid\sigma_{\eta_1}^2 &\sim \mathrm{N}_r\left( \bm{0}, \sigma_{\eta_1}^2 I_r \right), \nonumber\\ 
\phi_\beta,\phi_\eta &\overset{\mathrm{ind.}}{\sim} \mathrm{Unif}(-1,1), & \sigma_\beta^2 &\sim \mathrm{IG}(a_\beta,b_\beta), \label{eq:AL_lon_priors} \\
\sigma_{\eta_1}^2,\sigma_\eta^2 &\overset{\mathrm{ind.}}{\sim} \mathrm{IG}(a_\eta,b_\eta), & \sigma_t &\overset{\mathrm{ind.}}{\sim} \mathrm{IG}(a_\sigma,b_\sigma), \quad t=1,\ldots,T. \nonumber
 \end{align} 
Here, $\sigma_{\beta_1}^2$ is fixed, while $\sigma_\beta^2$ is the innovation variance of the regression-coefficient process. 
Similarly, $\sigma_{\eta_1}^2$ governs the initial spatial state and $\sigma_\eta^2$ is the innovation variance of the spatial process while $\phi_\beta$ and $\phi_\eta$ are the autoregressive parameters for the fixed and random effects, respectively.
The fixed hyperparameters satisfy $a_\beta,b_\beta,a_\eta,b_\eta,a_\sigma,b_\sigma>0$. 
For all AL and Laplace models, we set the fixed prior variance for the regression coefficients to 10.
Thus, $\sigma_\beta^2=10$ in AL-CS, while $\sigma_{\beta_1}^2=10$ in AL-T and Laplace-T.
All inverse-gamma hyperparameters are set to 0.1, so that $a_\sigma=b_\sigma=a_\eta=b_\eta=a_\beta=b_\beta=0.1$.
The MCMC and VB implementations use the same prior settings.
Conditional on the latent variables, most model components have standard full conditional distributions, resulting in an efficient Gibbs sampler.

For large datasets or when comparing a large number of candidate models, it may be impractical to run MCMC repeatedly.
The third component of our contribution is therefore a VB algorithm for approximate posterior inference under the cross-sectional model~(\ref{eq:AL_cs}) and the temporal model~(\ref{eq:AL_lon}) with priors~(\ref{eq:AL_lon_priors}).
Although VB has previously been developed for AL and dynamic exAL quantile regression \citep{waldmann_2015,Barata2022}, our computational contribution is the derivation of closed-form coordinate updates for the survey-weighted temporal small-area model.
VB provides a faster alternative to Gibbs sampling by replacing posterior simulation with optimization \citep{blei_2017}.
Rather than generating posterior draws through a Markov chain, VB approximates the posterior by a distribution $q(\bm\Theta)$ belonging to a tractable family $\mathcal{Q}$. 
The approximation is chosen to minimize the Kullback--Leibler (KL) divergence from $q(\bm\Theta)$ to the true posterior \[ q^*(\bm\Theta) = \mathop{\arg\min}_{q\in\mathcal{Q}} \operatorname{KL} \left\{ q(\bm\Theta) \,\|\, p(\bm\Theta\mid\bm y) \right\}, \] where $\bm\Theta$ denotes the collection of model parameters and latent variables.

Equivalently, minimizing the KL divergence is achieved by maximizing the evidence lower bound (ELBO),
\[
 \mathcal{L} = \mathbb{E}_q[\log p(\bm\Theta,\bm y)] - \mathbb{E}_q[\log q(\bm\Theta)].
\]
We take $\mathcal{Q}$ to be the {mean-field variational family}, which assumes the latent variables partition into mutually independent factors
\[
  q(\bm\Theta)= \prod_{j=1}^J q_j(\bm\Theta_j).
\]
This factorization yields closed-form coordinate updates for the variational factors.
Convergence is assessed by specifying a threshold $\varepsilon >0$ and terminating when
\[\left| \mathcal L^{(k)}-\mathcal L^{(k-1)} \right|<\varepsilon.\]
After convergence, the optimized variational factors define the approximate posterior $q^*(\bm\Theta)$ and we sample values of our parameters of interest (i.e., $\bm\beta$ and $\bm\eta$) from this variational distribution.
Further details of the Gibbs sampling and variational Bayes algorithms are provided in the Supplementary Material.

\subsection{Population estimates}
Given posterior draws from the fitted model, we propose two procedures for estimating finite population quantiles.
The choice of procedure depends on whether the desired analysis is to be carried out at the level of model predictors or at an aggregated level.

If analysis is carried out at the level of aggregation at which the model was fit (for example, covariates are race and age with a state-level random effect and we wish to produce estimates by race, age, and state) then the estimated linear predictor is exactly the conditional quantile.
That is, for posterior draw $j$, the estimated conditional $p$th quantile for subpopulation $k$ is
\[\widehat{Q}^{(j)}_{p,kt} = \bm x_{kt}'\bm\beta_t^{(j)} +\bm\psi_k'\bm\eta_t^{(j)}.\]
Here, $\bm{x}_{kt}$ denotes the covariate vector for subpopulation $k$ at time $t$ and $\bm{\psi}_k$ denotes the spatial basis vector for its area. 
A point estimate may be obtained from the posterior mean or median of $\{\widehat{Q}_{p,kt}^{(j)}\}_{j=1}^{J}$ for each $k$, and a $100(1-\alpha)\%$ credible interval may be obtained from its empirical $\alpha/2$ and $1-\alpha/2$ quantiles.

For target domains that aggregate over multiple demographic, geographic, or covariate profiles, conditional quantiles cannot generally be aggregated directly.
In particular, the $p$th quantile of a mixture of conditional distributions is not generally equal to the average, or the $p$th quantile, of the corresponding conditional quantiles.
We therefore estimate these finite-population quantities by simulating a population from the posterior predictive distribution and aggregating this synthetic population.
For posterior draw $j$, define the linear predictor for population unit $i$ at time $t$ as 
\begin{equation} 
\mu_{it}^{(j)}(p) = \bm{x}_{it}^{\prime}\bm{\beta}_{t}^{(j)}(p) + \bm{\psi}_{i}^{\prime}\bm{\eta}_{t}^{(j)}(p).
\label{eq:population_linear_predictor}
\end{equation}
Here, $j$ indexes either an MCMC posterior draw or a draw from the fitted variational approximation.
For each population unit, we generate 
\begin{equation} \widetilde y_{it}^{(j)} \mid \mu_{it}^{(j)}(p),\sigma_t^{(j)} \sim \mathrm{AL}_p \left( \mu_{it}^{(j)}(p), \sigma_t^{(j)} \right), \qquad i=1,\ldots,N_t,\quad t=1,\ldots,T.
\label{eq:pop_gen}
\end{equation} 
Here, unlike in conditional quantile estimation, we use the AL as a generative model.
For each posterior draw $j$ and target subpopulation $\mathcal U_{tk}$ we compute the empirical $p$th quantile of the generated values $\{\tilde{y}_{it}^{(j)}\}$. 
Repeating this procedure for $j=1,\ldots, J$ yields a posterior distribution of the finite-population quantile for the given subpopulation.
Point estimates and credible intervals are obtained from the corresponding posterior means or medians and empirical quantiles.
Because each generated population may be aggregated differently, the same draws can be used to estimate quantiles for multiple demographic and geographic domains.

\section{Empirical Simulation}\label{sec:sim}
To assess the proposed models, we carry out an empirical simulation study.
We treat the 2018--2025 SIPP data of householders in the contiguous United States over the age of $18$ as a finite population from which we draw repeated subsamples.
In this way, the original structure of the data is retained without specifying a parametric model to generate the finite population.
The resulting population consists of 155,818 responses from 83,159 households in 48 states, which form $8\times48\times9\times4=13,824$ year-area-age-race cells.
Of these cells, 9,001 are nonempty in the finite population used for the simulation (all 13,824 year-area-age-race cells are estimated in the data analysis in Section~\ref{sec:data_analysis}).
We aggregate the responses over age and sex to obtain the 1,422 nonempty year-area-race target domains used to evaluate the estimators.
In order to preserve the longitudinal structure of the data, if a household is sampled, then all of its responses are included in the sample.

We compare the proposed temporal AL model (\ref{eq:AL_lon}) (AL-T) to a direct estimator (DE), the longitudinal Gaussian unit-level model of \citet{Vedensky2025} (Gauss-T), and the cross-sectional AL model (\ref{eq:AL_cs}) (AL-CS).
Gauss-T accounts for within-household dependence by conditioning each follow-up response on the household's previous centered response, while its area effects evolve according to an AR(1) process.
AL-CS is fitted separately to each year of data. 
We also consider a temporal Laplace unit-level model (Laplace-T), which is obtained by fixing $p=0.5$ in the AL-T model.
This comparison assesses the benefit of targeting the quantile of interest rather than fitting only the conditional median.
All of these models are sampled with MCMC.
We also consider VB samplers for the AL-CS and AL-T models.

We include as covariates sex categories (male or female), race (Asian, Black, white, or other), and age (15-24, 25-34, 35-44, 45-54, 55-59, 60-64, 65-74, 75-84, or 85+).
These categorizations match the available Census population data.
We center and scale the response before model fitting and report the simulation results on this transformed scale for readability.

To make the samples informative, we sample in proportion to the size variable with Poisson sampling \citep{brewer_early_hanif_1984}
\[s_{i} = \log\left(1 + \exp\left\{2 +  \frac{\bar{y}_{i\cdot}^{*} + x_{i,age}}{2} \right\}\right),\]
where $\bar{y}_{i\cdot}$ is the mean of a household's responses averaged over all timepoints for which it provided a response and $\bar{y}_{i\cdot}^{*}$ is its standardized value across the 83,159 households,
and $x_{i,age}$ is the age category the householder belongs to represented as an integer, with $1$ indexing the lowest age category and $9$ the highest.
This makes older, wealthier respondents more likely to appear in the sample.
Samples were drawn with an expected sample size equal to 3\% of the population.

For each sample we consider estimates at the $0.10, 0.25, 0.50, 0.75,$ and $0.90$ quantiles.
All MCMC samplers are run for 2,000 iterations with the first 1,000 discarded for burn-in.
Visual inspection of traceplots suggests no lack of convergence.
For the VB samplers, we use $\varepsilon=10^{-4}$ and draw 1,000 samples from the converged variational distribution.
The direct estimator for each quantile is calculated using the \citet{Woodruff1952} method provided by the R \textit{survey} library \citep{lumley2010}.
The direct estimator variances are calculated using a logistic approximation, which prevents CI bounds on the estimated cdf from exceeding 0 or 1 \citep{Shah2006}.
The AL models are fit separately for each quantile.
Because the remaining models do not vary by quantile, they are fit once and separate posterior predictions are generated for each quantile level as follows.

To obtain model estimates for the population values, we sample values for the entire, known, population from the given model.
That is, if $\hat{\mu}_{it}^{(j)}=\bm x_{it}'\hat{\bm\beta}_t^{(j)} + \bm\psi_i'\hat{\bm\eta}_t^{(j)}$ is the estimated linear predictor at MCMC iteration $j$ for the AL-T model, we generate
\begin{equation}
    \hat{y}_{it}^{(j)} \mid \hat{\mu}_{it}^{(j)}, \hat{\sigma}_t^{(j)} \sim AL_p(\hat{\mu}_{it}^{(j)}, \hat{\sigma}_t^{(j)}) \label{eq:sim_pop_gen}
\end{equation}
for each $t=1,\ldots, T$ and $i=1,\ldots,N_t$.
Here, $\hat{\sigma}_t^{(j)}$ denotes the $j$th posterior draw of the AL scale for year $t$.
For the Gaussian model, we instead generate $\hat{y}_{it}^{(j)}\mid\hat{\mu}_{it}^{(j)},\hat{\sigma}^{2(j)}\sim \mathrm{N}(\hat{\mu}_{it}^{(j)},\hat{\sigma}^{2(j)})$, where $\hat{\sigma}^{2(j)}$ denotes the $j$th posterior draw of the Gaussian conditional variance.
We then aggregate the estimates by race, state, and year and calculate the $p$th empirical quantile for each race-state combination at each iteration.
Lastly, we compute posterior means and credible intervals (CIs) for each target domain, which allow us to calculate root mean squared error (RMSE), 95\% CI coverage rate, mean absolute bias (MAB), and the interval score \citep{Gneiting_2007}.
For the $D$ evaluated domains, let $Q_d$ denote the finite-population truth in domain $d$ and let $\widehat Q_{rd}$ denote its point estimate in replicate $r$.
We define
\[\operatorname{RMSE}=\left(\frac{1}{D}\sum_{d=1}^D\left(\frac{1}{|\mathcal R_d|}\sum_{r\in\mathcal R_d}(\widehat Q_{rd}-Q_d)^2\right)\right)^{1/2}\]
and
\[
\operatorname{MAB}=\frac{1}{D}\sum_{d=1}^D\left|\frac{1}{|\mathcal R_d|}\sum_{r\in\mathcal R_d}(\widehat Q_{rd}-Q_d)\right|,
\]
where $\mathcal R_d$ is the set of replicates available for domain $d$.
The interval score is
\[IS_\alpha(\ell, u; x) = (u-\ell)+ \frac{2}{\alpha}(\ell - x)\mathbf{1}\{x<\ell\} + \frac{2}{\alpha}(x-u)\mathbf{1}\{x>u\}.\]
This last measure penalizes overly wide predictive intervals, so that an estimator with narrower intervals containing the true value yields a lower score.

The simulation results are summarized in Tables~\ref{tab:simulation_common_rmse_bias}--\ref{tab:simulation_full_is_coverage}.
Tables~\ref{tab:simulation_common_rmse_bias} and~\ref{tab:simulation_full_rmse_bias} report RMSE with MAB in parentheses, while Tables~\ref{tab:simulation_common_is_coverage} and~\ref{tab:simulation_full_is_coverage} report interval score, with empirical coverage rate of the corresponding 95\% intervals in parentheses.
Within each pair of tables, the first table restricts the comparison to the set of domains for which a DE point estimate and interval could be produced and reports results for the DE alongside the model-based estimators.
The second table includes all 1,422 target domains in each replicate and therefore compares only the model-based estimators as these are able to make predictions for every domain, whereas the DE cannot produce a point estimate or interval due to low sample size.
Because the subset of domains for which the DE is available varies across quantile level, comparisons in Tables~\ref{tab:simulation_common_rmse_bias} and~\ref{tab:simulation_common_is_coverage} should be made within each column since averages are taken over different numbers.
Moreover, the variability and scale of the target domains increase with the quantile level as well, so that estimation is more difficult in the upper tail in general, and the scale of our reported metrics increases with quantile level.

 \begin{table}[htbp]
\centering
\small
\caption{RMSE and MAB on domains with an available DE. 
Entries are RMSE (MAB) on the standardized-response scale. 
Lower RMSE and MAB are better. 
Bold identifies the lowest RMSE and MAB at each quantile. 
Within each represented domain, calculations use only the replicates in which the DE point estimate and interval are available.
The numbers of unique domains at $p=0.10$, $0.25$, $0.50$, $0.75$, and $0.90$ are 902, 902, 902, 851, 813, respectively.
Across 100 replicates, these comprise 43,275, 43,275, 43,268, 37,020, 22,082 domain-replicate evaluations, respectively.}
\label{tab:simulation_common_rmse_bias}
\resizebox{\linewidth}{!}{%
\begin{tabular}{lrrrrr}
\toprule
Method & $p=0.10$ & $p=0.25$ & $p=0.50$ & $p=0.75$ & $p=0.90$ \\
\midrule
AL-T (MCMC) & \textbf{0.017} (0.0084) & 0.020 (0.0126) & 0.043 (0.0294) & 0.108 (0.0746) & 0.298 (0.2240) \\
AL-CS (MCMC) & 0.019 (0.0098) & 0.023 (0.0135) & 0.049 (0.0314) & 0.118 (0.0816) & 0.347 (0.2612) \\
AL-T (VB) & \textbf{0.017} (\textbf{0.0076}) & \textbf{0.019} (\textbf{0.0120}) & \textbf{0.041} (\textbf{0.0283}) & \textbf{0.106} (\textbf{0.0735}) & \textbf{0.288} (0.2182) \\
AL-CS (VB) & 0.018 (0.0084) & 0.022 (0.0131) & 0.048 (0.0300) & 0.115 (0.0784) & 0.341 (0.2544) \\
Laplace-T (MCMC) & 0.290 (0.2819) & 0.105 (0.0972) & 0.043 (0.0295) & 0.127 (0.1018) & 0.292 (0.2236) \\
Gauss-T & 0.802 (0.7822) & 0.357 (0.3361) & 0.166 (0.1448) & 0.564 (0.5499) & 0.806 (0.7564) \\
\midrule
DE & 0.326 (0.1052) & 0.321 (0.0960) & 0.310 (0.0774) & 0.409 (0.0971) & 0.874 (\textbf{0.2070}) \\
\bottomrule
\end{tabular}
}
\end{table}

\begin{table}[htbp]
\centering
\small
\caption{RMSE and MAB on the full set of domains. 
Entries are RMSE (MAB) on the standardized-response scale. 
Lower RMSE and MAB are better. 
Bold identifies the lowest RMSE and MAB at each quantile. 
DE is excluded because it is not available on the full set of domains.
All model-based estimators are evaluated on all 1,422 target domains in each of 100 replicates (142,200 domain-replicate evaluations per quantile).}
\label{tab:simulation_full_rmse_bias}
\resizebox{\linewidth}{!}{%
\begin{tabular}{lrrrrr}
\toprule
Method & $p=0.10$ & $p=0.25$ & $p=0.50$ & $p=0.75$ & $p=0.90$ \\
\midrule
AL-T (MCMC) & \textbf{0.073} (0.0325) & \textbf{0.069} (0.0331) & 0.087 (0.0449) & 0.179 (0.1044) & 0.948 (0.3019) \\
AL-CS (MCMC) & 0.074 (\textbf{0.0323}) & 0.071 (0.0337) & 0.092 (0.0473) & 0.191 (0.1107) & 0.966 (0.3313) \\
AL-T (VB) & 0.075 (0.0334) & \textbf{0.069} (\textbf{0.0330}) & \textbf{0.086} (\textbf{0.0437}) & 0.180 (\textbf{0.1042}) & 0.948 (0.2992) \\
AL-CS (VB) & 0.076 (0.0335) & 0.071 (0.0340) & 0.091 (0.0456) & 0.192 (0.1085) & 0.968 (0.3252) \\
Laplace-T (MCMC) & 0.266 (0.2477) & 0.117 (0.0971) & 0.087 (0.0449) & \textbf{0.171} (0.1206) & \textbf{0.928} (\textbf{0.2667}) \\
Gauss-T & 0.714 (0.6637) & 0.329 (0.2979) & 0.175 (0.1450) & 0.526 (0.4940) & 1.129 (0.7103) \\
\bottomrule
\end{tabular}
}
\end{table}

\begin{table}[htbp]
\centering
\small
\caption{Interval score and coverage on domains with an available DE.
Entries are 95\% interval score (coverage percentage). 
Lower interval scores are better, and coverage closer to the nominal 95\% level is better.
Bold identifies the lowest interval score at each quantile. 
Within each represented domain, calculations use only the replicates in which the DE point estimate and interval are available.
The numbers of unique domains at $p=0.10$, $0.25$, $0.50$, $0.75$, and $0.90$ are 902, 902, 902, 851, 813, respectively.
Across 100 replicates, these comprise 43,275, 43,275, 43,268, 37,020, 22,082 domain-replicate evaluations, respectively.}
\label{tab:simulation_common_is_coverage}
\resizebox{\linewidth}{!}{%
\begin{tabular}{lrrrrr}
\toprule
Method & $p=0.10$ & $p=0.25$ & $p=0.50$ & $p=0.75$ & $p=0.90$ \\
\midrule
AL-T (MCMC) & 0.108 (99.6\%) & 0.142 (99.7\%) & \textbf{0.234} (97.6\%) & \textbf{0.561} (87.1\%) & \textbf{2.651} (60.3\%) \\
AL-CS (MCMC) & 0.118 (99.6\%) & 0.156 (99.7\%) & 0.258 (98.0\%) & 0.566 (89.1\%) & 2.906 (61.4\%) \\
AL-T (VB) & \textbf{0.095} (99.1\%) & \textbf{0.130} (98.9\%) & \textbf{0.234} (94.2\%) & 0.707 (76.6\%) & 3.624 (45.7\%) \\
AL-CS (VB) & 0.101 (98.4\%) & 0.140 (98.3\%) & 0.258 (94.2\%) & 0.736 (77.3\%) & 4.398 (45.2\%) \\
Laplace-T (MCMC) & 5.227 (14.6\%) & 0.814 (57.6\%) & \textbf{0.234} (97.7\%) & 1.045 (57.5\%) & 4.019 (41.2\%) \\
Gauss-T & 16.860 (8.7\%) & 3.464 (45.3\%) & 0.993 (82.8\%) & 10.056 (15.0\%) & 17.176 (11.4\%) \\
\midrule
DE & 3.389 (55.4\%) & 3.183 (59.7\%) & 2.939 (58.9\%) & 4.282 (49.6\%) & 10.928 (28.4\%) \\
\bottomrule
\end{tabular}
}
\end{table}

\begin{table}[htbp]
\centering
\small
\caption{Interval score and coverage on the full set of domains. 
Entries are 95\% interval score (coverage percentage). 
Lower interval scores are better, and coverage closer to the nominal 95\% level is better.
Bold identifies the lowest interval score at each quantile. 
DE is excluded because it is not available on the full set of domains.
All model-based estimators are evaluated on all 1,422 target domains in each of 100 replicates (142,200 domain-replicate evaluations per quantile).}
\label{tab:simulation_full_is_coverage}
\resizebox{\linewidth}{!}{%
\begin{tabular}{lrrrrr}
\toprule
Method & $p=0.10$ & $p=0.25$ & $p=0.50$ & $p=0.75$ & $p=0.90$ \\
\midrule
AL-T (MCMC) & 0.248 (99.5\%) & 0.318 (99.5\%) & 0.491 (97.8\%) & \textbf{1.040} (90.0\%) & \textbf{4.288} (71.5\%) \\
AL-CS (MCMC) & 0.261 (99.5\%) & 0.333 (99.5\%) & 0.514 (98.3\%) & 1.057 (91.5\%) & 4.434 (72.2\%) \\
AL-T (VB) & \textbf{0.237} (99.0\%) & \textbf{0.309} (99.0\%) & \textbf{0.490} (95.6\%) & 1.140 (83.5\%) & 4.970 (62.3\%) \\
AL-CS (VB) & 0.243 (98.5\%) & 0.318 (98.6\%) & 0.508 (95.6\%) & 1.195 (83.8\%) & 5.581 (61.0\%) \\
Laplace-T (MCMC) & 3.721 (42.3\%) & 0.852 (72.9\%) & 0.491 (97.9\%) & 1.164 (72.2\%) & 4.817 (57.5\%) \\
Gauss-T & 11.692 (37.1\%) & 2.937 (65.3\%) & 1.354 (89.1\%) & 6.990 (45.7\%) & 12.870 (35.8\%) \\
\bottomrule
\end{tabular}
}
\end{table}

When compared to the direct estimator (Tables~\ref{tab:simulation_common_rmse_bias} and~\ref{tab:simulation_common_is_coverage}) all four AL estimators obtain substantially lower RMSE and IS than the DE.
On the common set of estimable domains, AL-T (VB) has the lowest RMSE at every quantile level and the lowest MAB through $p=0.75$.
At $p=0.90$, the DE has slightly lower MAB, but its RMSE is more than three times as large as that of AL-T (VB).
The temporal AL models also obtain lower RMSE and IS than their cross-sectional counterparts (that is, AL-T (VB) outperforms AL-CS (VB) and AL-T (MCMC) outperforms AL-CS (MCMC)) at every quantile level on the subset of domains for which a DE interval is available.
On the full set of 1,422 target domains (Tables~\ref{tab:simulation_full_rmse_bias} and~\ref{tab:simulation_full_is_coverage}), both temporal AL estimators obtain lower RMSE and interval score than their corresponding cross-sectional estimators at every quantile level.
On this full set, AL-T (VB) and AL-T (MCMC) have nearly identical RMSE and MAB, while VB is substantially faster to fit.
As shown in Table~\ref{tab:simulation_runtime_revised}, replacing MCMC with VB reduces the median fitting time for the temporal model from 11.56 to 1.28 minutes.
\begin{table}[htbp]
\centering
\small
\caption{Median wall-clock runtime per replicate across all five target quantiles (100 replicates).
Fit, prediction, and total times are separate medians of the per-replicate sums across the five quantiles and therefore need not add exactly.
Laplace-T and Gauss-T are omitted because they are fitted once, whereas the other models are fitted separately at each quantile, so their runtimes are not directly comparable.
All four estimators were run on the same 100 samples.
Computation used a Dell C6525 with two 64-core AMD EPYC 7713 processors and 512~GB DDR4 memory.}
\label{tab:simulation_runtime_revised}
\begin{tabular}{lrrr}
\toprule
Method & Fit (min) & Prediction (min) & Total (min) \\
\midrule
AL-T (MCMC) & 11.56 & 5.65 & 17.26 \\
AL-CS (MCMC) & 7.70 & 5.61 & 13.32 \\
AL-T (VB) & 1.28 & 5.76 & 7.10 \\
AL-CS (VB) & 0.16 & 5.79 & 5.94 \\
\bottomrule
\end{tabular}
\end{table}

The AL estimators obtain substantially lower interval scores than the DE at every quantile, indicating a better tradeoff between interval width and failures to contain the finite-population truth.
Their empirical coverage is conservative at lower quantiles and falls below the nominal 95\% level in the upper tail, however, particularly for VB.
Among the AL estimators, MCMC provides better IS in the upper tail.

The Gaussian model performs worst among the model-based estimators and generally underperforms even the DE. 
The heavy tails of the data inflate the Gaussian variance so that it ends up underestimating the low quantiles and overestimating the upper quantiles.
The Laplace model agrees with AL at the median, as expected, but its performance deteriorates away from the median.
Its interval score exceeds that of the DE at $p=0.10$, while its interval coverage remains unreliable in the tails.
Although it obtains competitive RMSE at some upper quantiles, including the lowest RMSE at $p=0.75$ and $p=0.90$ in the model-based comparison, its interval estimates are not reliable in the tails.

\section{Data Analysis}\label{sec:data_analysis}
We perform an analysis of the full 2018--2025 SIPP data, which records household wealth for reference years 2017--2024.
To demonstrate our model's applicability to small area estimation, we consider a demographic breakdown of race and age by state and year.
As population information for post-stratification, we use the 2020 Census Demographic and Housing Characteristics File tables H13A--H13G, which we refer to collectively as H13.\footnote{\url{https://data.census.gov/table/DECENNIALDHC2020.H13A?t=Populations+and+People&y=2020&d=DEC+Demographic+and+Housing+Characteristics}}
These tables provide household counts cross-classified by state, age, and race.
Because comparable annual census counts with the H13 cross-classification are unavailable, we use the 2020 H13 counts for every reference year.
State-level estimates are therefore standardized to each state's 2020 age-by-race composition.
To match the classification used in SIPP, we collapse the H13 categories for American Indian and Alaska Native, Native Hawaiian and Other Pacific Islander, some other race, and two or more races into a single category.
The resulting analysis contains 13,824 domains.
Of these, 4,823 do not appear in the SIPP sample and only 15.6\% have a sample size greater than 20.
For domains that do not appear in the SIPP sample, predictions are obtained from their age and race covariates and state-year spatial effect, with the associated parameters estimated from the observed domains.

We fit the temporal AL model for three different quantile levels: $0.10$, $0.50$, and $0.90$.
We refer to the three models as $\text{AL}_{10},$ $\text{AL}_{50}$, and $\text{AL}_{90},$ respectively.
When modeling multiple quantiles, a known shortcoming of ``multiple QR" models \citep{CastilloMateo2023} such as ours is that the different quantiles may ``cross", violating the monotonicity of the quantile function.
To avoid this pitfall, we present the different models individually to assess model performance at different levels and do not compare quantiles directly.

\begin{figure}[ht!]
    \centering
     \providecommand{\areamedianfigureheight}{\textheight}
    \includegraphics[width=\linewidth,height=\areamedianfigureheight,keepaspectratio]{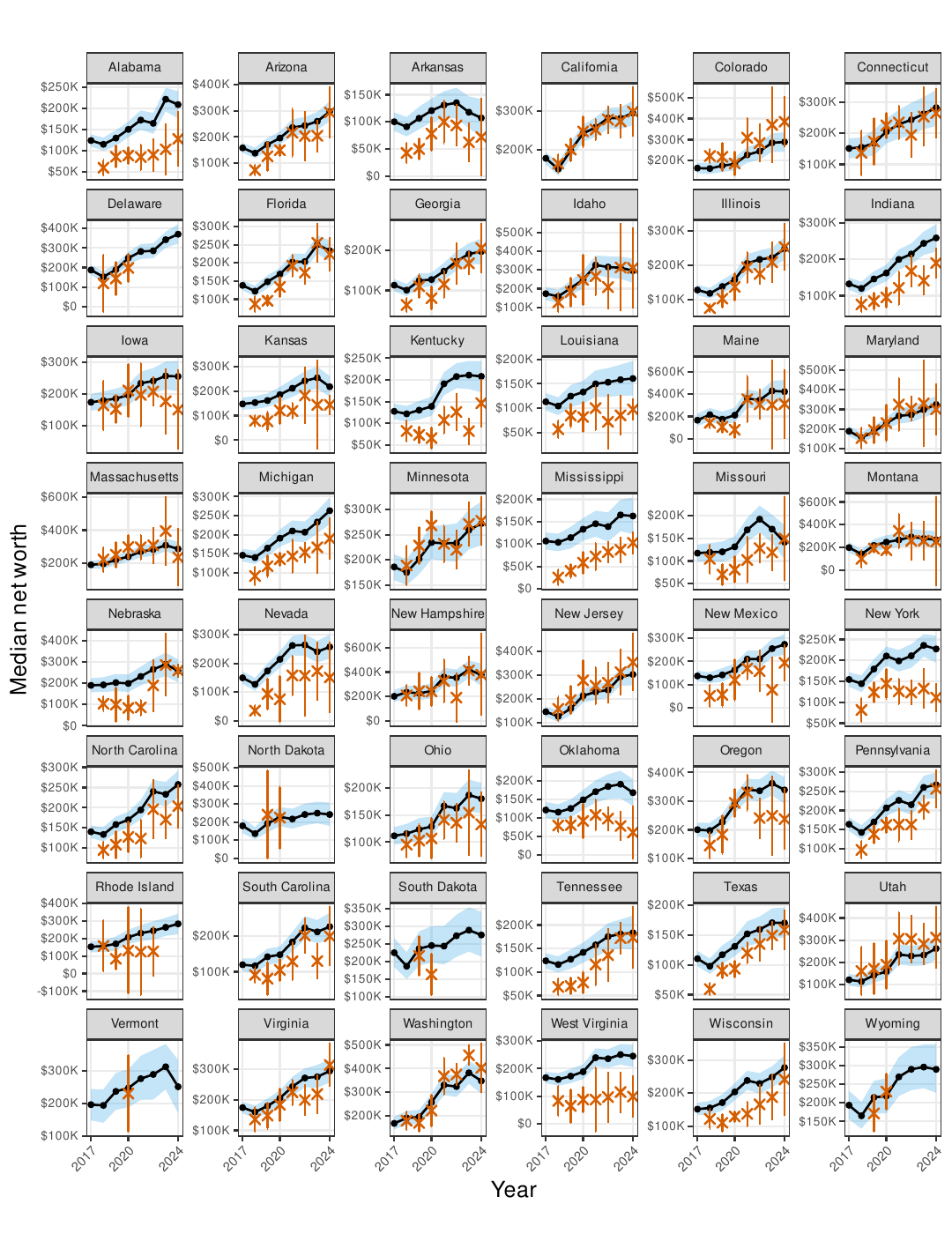}
        \caption{Time series plot of area-level model point estimates and 95\% posterior bands for the median produced by generating a synthetic population. 
        Crosses at each time point indicate the published SIPP direct estimates with 95\% confidence intervals.
        Areas with low sample sizes lack a reported value due to disclosure limitations.
    \label{fig:area_medians}}
\end{figure}

We use race and age as covariates in the model.
We exclude sex because it is not included in the Census population table used for poststratification.
Moreover, differences in household wealth by the sex of the householder cannot generally be interpreted as evidence of individual-level gender wealth gaps.
A more appropriate analysis would require comparing single-person households led by women and those led by men \citep{balestra_etal_2025}.
For each model, we run four independently initialized chains for 8,000 iterations and discard the first 1,000 iterations as burn in.
From each chain, we retain one draw every seven post-burn-in iterations and generate posterior predictions, giving 1,000 draws per chain and 4,000 posterior predictions for each model.
MCMC convergence diagnostics for the reported finite-population domain estimates from all three quantile models are provided in the supplementary material.
At each retained draw, we generate one response for every household represented by the H13 population counts and calculate empirical quantiles within each target domain.
The same generated population can be aggregated to obtain estimates at coarser resolutions, including the state-level estimates considered below.
All reported model estimates pool the 4,000 generated populations across the four chains.
%
To illustrate, we produce a set of area-level estimates from the $\text{AL}_{50}$ model via this method, aggregating the generated household responses over age and race within each state and year and calculating the empirical median.
Figure~\ref{fig:area_medians} displays the resulting estimates of median net worth for each state and each year along with 95\% posterior CIs.
Crosses with error bars mark direct estimates published by the Census Bureau and their 95\% confidence intervals.\footnote{\url{https://www.census.gov/topics/income-poverty/wealth/data/tables.html}}
 These values are produced independently for each state and each year.
The model also provides estimates for state-year combinations with no published direct estimate.
Vermont provides the clearest example, with only one published median estimate over the study period but model estimates and posterior intervals for all eight years.
The model estimates generally overlap with the reported confidence intervals, but with a smoother temporal trend. 
The model CIs are generally narrower than the DE CIs.
Disagreement between the model and published direct estimates should not necessarily be interpreted as model error because, as noted in Section~\ref{sec:methods}, the direct quantile estimator is not generally design unbiased.
In the simulation study, where the finite-population values are known, the direct estimator of the median has substantially higher RMSE and MAB than the temporal AL estimators, along with poor interval coverage.
Differences such as those in Figure~\ref{fig:area_medians} are therefore not unexpected, particularly for state-year combinations with small sample sizes, where the published direct estimates are themselves subject to substantial sampling error and finite-sample bias.

As a point of comparison, we also produce direct estimates for the three quantile levels using the same settings as in Section~\ref{sec:sim} but with variance estimates now produced using replicate weights \citep{lumley2010}.
Because of the many cells with low sample sizes, it is only possible to produce direct estimates for 59.5\% of the domains.
Direct estimate standard errors and confidence intervals can only be produced for 45.2\%, 41.1\%, and 20.9\% of domains at $p=0.10$, $p=0.50$, and $p=0.90$, respectively.

\begin{figure}[ht!]
    \centering
    \includegraphics[width=\linewidth]{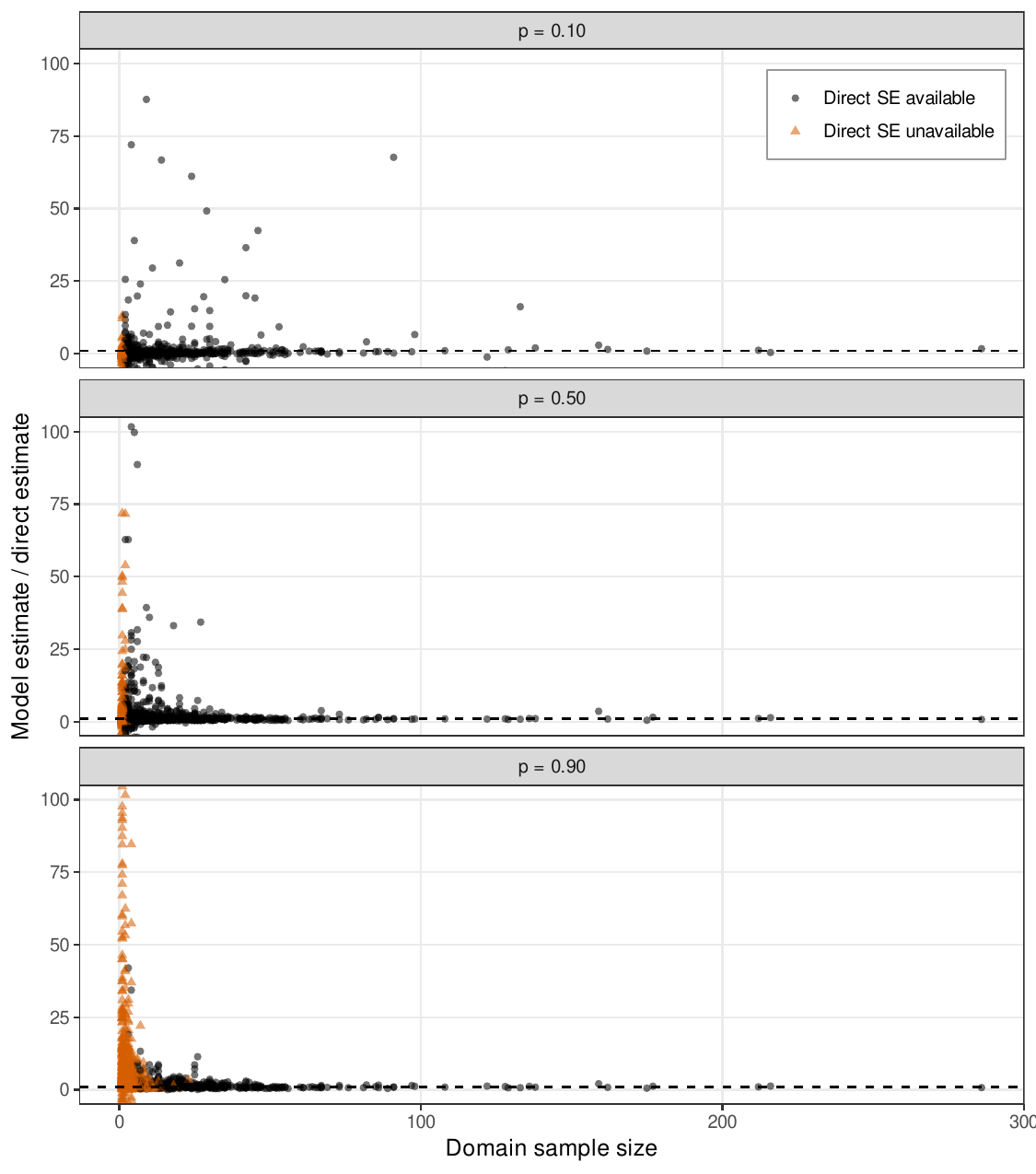}
        \caption{Ratio of model estimates to direct estimates by cell sample size for the $10$th, $50$th, and $90$th quantile model.
        The vertical axis is truncated at 100 for readability.
        Estimates for which a direct estimate standard error could not be produced due to low domain size are marked as triangles. 
    \label{fig:estimate_ratio}}
\end{figure}

\begin{figure}[ht!]
    \centering
    \includegraphics[width=\linewidth]{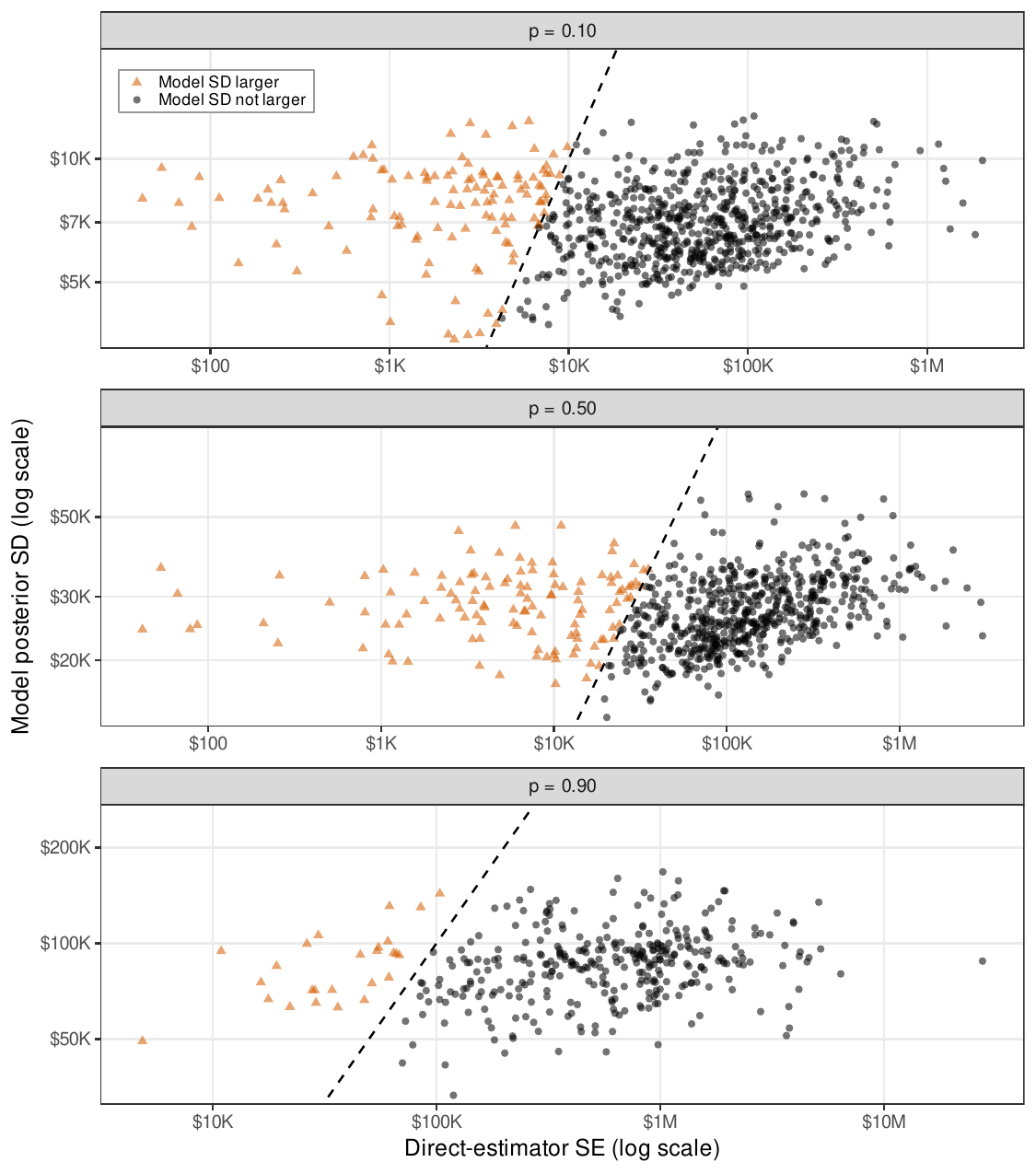}
        \caption{Scatterplots of model posterior standard deviations against direct estimate standard errors for the $10$th, $50$th, and $90$th quantile models on log scales. 
        Points to the right of the $x=y$ line indicate domains for which the direct estimate has higher standard errors.
       The model posterior standard deviation is no larger for $84.8\%$, $83.3\%$, and $92.8\%$ of the domains at $p=0.10$, $p=0.50$, and $p=0.90$, respectively.
    \label{fig:SE_ratio}}
\end{figure}

Figure~\ref{fig:estimate_ratio} shows the ratio of model estimates to direct estimates by domain sample size for all three quantiles.
While the two estimators agree on cells with large sample sizes, the variation for smaller cells leads to large discrepancies, where the model provides more smoothing.
Because direct estimates become more precise as domain sample size increases, this pattern suggests that the model's borrowing of strength has its greatest effect in domains where direct estimation is least reliable, while the model-based and direct estimates increasingly agree in larger domains.
Figure~\ref{fig:SE_ratio} plots the model posterior standard deviations against the direct estimate standard errors for all three quantiles.
The model posterior standard deviation is no larger than the direct estimate standard error for more than 80\% of the comparable domains at every quantile, with the largest proportion occurring at $p=0.90$.
The variability in the direct estimates and the high number of missing estimates make it unwieldy to plot estimates over space or time.
Meanwhile, the model, by borrowing strength across space and time, allows for the visualization of such trends.

\begin{figure}
    \includegraphics[width=\linewidth]{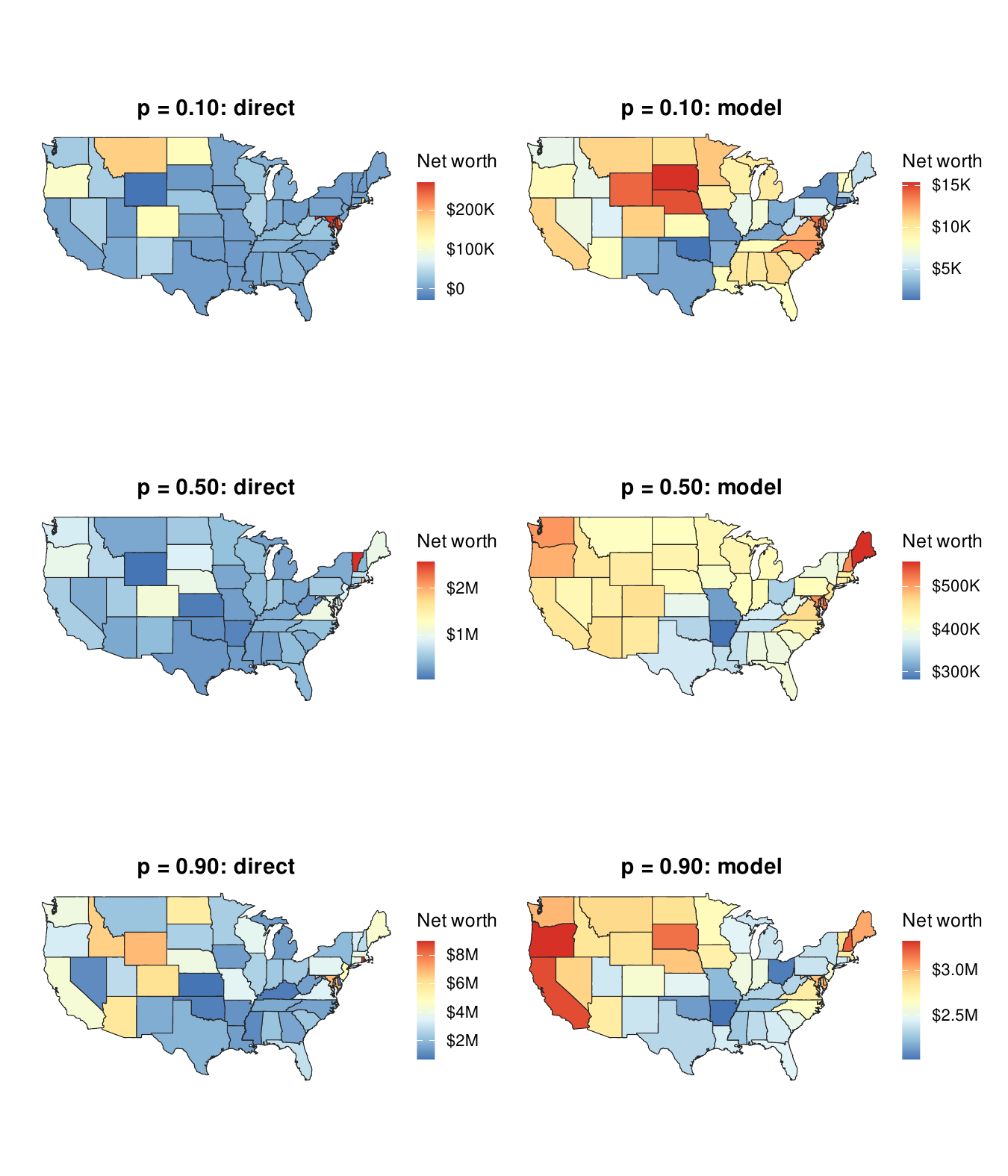}
        \caption{Areal map of direct estimates (left) and model estimates (right) for $p=0.10$, $p=0.50$, and $p=0.90$ for white householders aged 65 to 75 in reference year 2024.
        The direct and model panels use separate color scales.}
        \label{fig:areal}
\end{figure}

As another point of comparison, we plot direct estimates and model estimates of all three quantiles for white householders between the ages of 65 and 75 by state in the final survey reference year in Figure~\ref{fig:areal}.
The spatial pattern of the model estimates is generally smoother than the direct estimates, whose range is largely implausible.
For instance, the estimate for $p=0.10$ in Maryland is nearly \$270,000, which exceeds several of the estimates for median wealth in the middle plot.
In contrast, the model estimates for $p=0.10$ do not exceed \$16,000.
At $p=0.50$, six direct estimates exceed \$1 million and the maximum is \$2.56 million in Vermont, while the maximum model estimate is approximately \$560,000.
The contrast is also apparent at $p=0.90$, where the direct estimates reach \$8.94 million in Rhode Island while the model estimates range from \$2.02 million to \$3.31 million.

\section{Discussion}\label{sec:conclusion}
This paper develops a Bayesian unit-level framework for estimating domain quantiles of a finite population under informative sampling while borrowing strength across space and time.
With SIPP as a motivating example we show how such a model can be applied to wealth data, which are skewed and heavy-tailed.
The simulation results show that modeling quantiles directly and borrowing information over time improve estimation relative to Gaussian and cross-sectional alternatives.

The SIPP application further demonstrates the practical value of producing estimates across the wealth distribution when direct estimates are unavailable or highly variable.
By estimating domains with little or no sample support and smoothing unstable year-to-year variation, the model provides a more complete account of geographic and demographic differences in wealth than direct estimation alone.
The similar point estimates obtained by MCMC and VB, together with the shorter fitting time of VB, offer complementary options for applications with different computational requirements.
The framework may therefore also be useful for other skewed finite-population outcomes where interest extends beyond the mean.

Although we have only considered public use microdata in our data analysis, future work should explore finer geographies, perhaps through data linkage with a larger survey like the American Community Survey as \citet{Chenevert2017} do for a cross-sectional analysis.
In order to answer questions that require examining the relationship between quantiles, such as changes in wealth inequality over time, it would also be important to consider a joint model for quantiles that is not prone to crossing.
Future work should also investigate variance corrections for the AL pseudo-posterior.

\clearpage

\section*{Acknowledgments}
We thank Robert O. Munk and Shomik Ghosh for comments that helped improve an earlier version of this manuscript.

\section*{Funding}
This research was supported in part by the Census Bureau Dissertation Fellowship program, the Natural Sciences and Engineering Research Council of Canada, and partially supported by the NSF under NSF grants NCSE-2215168 and NCSE-2215169.

\bibliographystyle{apalike}
\bibliography{bibliography}
\clearpage

\appendix
\section{Full conditional distributions and variational Bayes algorithms}\label{sec:appendix}

\subsection{Asymmetric Laplace mixture representation}

Let $p$ denote the modeled quantile and define
\[
\theta=\frac{1-2p}{p(1-p)}
\qquad\text{and}\qquad
\gamma^2=\frac{2}{p(1-p)}.
\]
For observation $i$ in year $t$, the asymmetric Laplace working model has the
location-scale mixture representation
\begin{align*}
y_{it}\mid\nu_{it},\bm\beta_t,\bm\eta_t,\sigma_t
&\sim N\left(\mu_{it}+\theta\nu_{it},
\frac{\gamma^2\sigma_t\nu_{it}}{w_{it}}\right),\\
\nu_{it}\mid\sigma_t
&\sim \operatorname{Exp}\left(\frac{\sigma_t}{w_{it}}\right),
\end{align*}
where $\mu_{it}=\bm x_{it}'\bm\beta_t+\bm\psi_{it}'\bm\eta_t$ and
$\operatorname{Exp}(m)$ denotes an exponential distribution with mean $m$.
The weights $w_{it}$ below are the scaled survey weights
$\widetilde w_{it}$ defined in the main text.
The cross-sectional model is fitted separately within each year, with its own
$\bm\beta$ and $\bm\eta$. The temporal model uses
\[
\begin{aligned}
\bm\beta_1&\sim N(\bm 0,\sigma_{\beta_1}^2I_q), &
\bm\beta_t\mid\bm\beta_{t-1},\phi_\beta,\sigma_\beta^2
&\sim N(\phi_\beta\bm\beta_{t-1},\sigma_\beta^2I_q),\\
\bm\eta_1&\sim N(\bm 0,\sigma_{\eta_1}^2I_r), &
\bm\eta_t\mid\bm\eta_{t-1},\phi_\eta,\sigma_\eta^2
&\sim N(\phi_\eta\bm\eta_{t-1},\sigma_\eta^2I_r),
\end{aligned}
\]
for $t=2,\ldots,T$. Here $q$ is the number of fixed-effect coefficients and
$r$ is the number of retained basis functions. The initial fixed-effect
variance $\sigma_{\beta_1}^2$ is fixed. The remaining variance components
have inverse-gamma priors and $\phi_\beta$ and $\phi_\eta$ have uniform priors
on $(-1,1)$.

\subsection{Gibbs sampling}

Write $\operatorname{IG}(a,b)$ for an inverse-gamma distribution
with density
\[
f(z\mid a,b)=\frac{b^a}{\Gamma(a)}z^{-(a+1)}\exp(-b/z),
\qquad z>0.
\]
Thus $a$ is the shape and $b$ is the scale. The latent variable
$\nu_{it}$ has the generalized inverse Gaussian full conditional
\[
\nu_{it}\mid\cdot
\sim \operatorname{GIG}\left(\frac12,\chi_{it},\psi_{it}^{\nu}\right),
\]
where
$\chi_{it}=w_{it}(y_{it}-\mu_{it})^2/(\gamma^2\sigma_t)$ and
$\psi_{it}^{\nu}=w_{it}\{\theta^2/\gamma^2+2\}/\sigma_t$.
The scale parameter $\sigma_t$ has full conditional
\[
\sigma_t\mid\cdot
\sim \operatorname{IG}\left(a_\sigma+\frac{3n_t}{2},\;
b_\sigma+\sum_{i\in\mathcal S_t}w_{it}\nu_{it}
+\frac{1}{2\gamma^2}
\sum_{i\in\mathcal S_t}\frac{w_{it}}{\nu_{it}}
(y_{it}-\mu_{it}-\theta\nu_{it})^2\right).
\]

For the cross-sectional model, let
$\bm\xi=(\bm\beta',\bm\eta')'$,
$C=[X,\Psi]$, and
\[
D=\diag\left\{\frac{w_i}{\gamma^2\sigma\nu_i}\right\}.
\]
Then $\bm\xi\mid\cdot\sim N(\bm m_\xi,V_\xi)$, where
\begin{align*}
V_\xi&=\left(C'DC+\operatorname{blockdiag}
\{\sigma_{\beta}^{-2}I_q,\sigma_{\eta}^{-2}I_r\}\right)^{-1},\\
\bm m_\xi&=V_\xi C'D(\bm y-\theta\bm\nu).
\end{align*}
The full conditional for $\sigma_\eta^2$ is
\[
\sigma_\eta^2\mid\cdot\sim
\operatorname{IG}\left(a_\eta+\frac r2,
b_\eta+\frac12\bm\eta'\bm\eta\right).
\]

The full conditional updates can be expressed once for either
$\bm h_t=\bm\beta_t$ or $\bm h_t=\bm\eta_t$. Let $A_t=X_t$ for the
fixed effects and $A_t=\Psi_t$ for the spatial effects. Let $d_h$ be the
dimension of $\bm h_t$, let $v_{h1}$ be its initial variance, and let $v_h$
and $\phi_h$ be its innovation variance and autoregressive coefficient. With
\[
D_t=\diag\left\{\frac{w_{it}}{\gamma^2\sigma_t\nu_{it}}
\mathrel{:}i\in\mathcal S_t\right\},
\]
the full conditional is $\bm h_t\mid\cdot\sim
N(\bm m_{h,t},V_{h,t})$, where
\begin{align*}
V_{h,t}&=(A_t'D_tA_t+\lambda_{h,t}I_{d_h})^{-1},\\
\bm m_{h,t}&=V_{h,t}\{A_t'D_t\bm e_{h,t}+\bm b_{h,t}\},
\end{align*}
where $\bm e_{\beta,t}=\bm y_t-\Psi_t\bm\eta_t-\theta\bm\nu_t$ and
$\bm e_{\eta,t}=\bm y_t-X_t\bm\beta_t-\theta\bm\nu_t$. The quantities
contributed by the AR(1) prior are
\[
\begin{aligned}
\lambda_{h,1}&=v_{h1}^{-1}+\phi_h^2v_h^{-1}, &
\bm b_{h,1}&=\phi_hv_h^{-1}\bm h_2,\\
\lambda_{h,t}&=(1+\phi_h^2)v_h^{-1}, &
\bm b_{h,t}&=\phi_hv_h^{-1}(\bm h_{t-1}+\bm h_{t+1}),
&&2\leq t<T,\\
\lambda_{h,T}&=v_h^{-1}, &
\bm b_{h,T}&=\phi_hv_h^{-1}\bm h_{T-1}.
\end{aligned}
\]
The full conditional for $\phi_h$ is truncated normal
\[
\phi_h\mid\cdot\sim
N_{(-1,1)}\left(\frac{R_h}{S_h},\frac{v_h}{S_h}\right),
\]
where $S_h=\sum_{t=2}^T\bm h_{t-1}'\bm h_{t-1}$ and
$R_h=\sum_{t=2}^T\bm h_t'\bm h_{t-1}$.
The full conditional for $v_h$ is
\[
v_h\mid\cdot\sim
\operatorname{IG}\left(a_h+\frac{d_h(T-1)}2,
b_h+\frac12\sum_{t=2}^T
\|\bm h_t-\phi_h\bm h_{t-1}\|^2\right).
\]
The full conditional for $\sigma_{\eta_1}^2$ is
\[
\sigma_{\eta_1}^2\mid\cdot\sim
\operatorname{IG}\left(a_{\eta}+\frac r2,
b_{\eta}+\frac12\bm\eta_1'\bm\eta_1\right).
\]

\subsection{Variational Bayes algorithms}

The cross-sectional approximation is
\[
q(\bm\xi)q(\sigma_\eta^2)q(\sigma)
\prod_{i=1}^nq(\nu_i).
\]
The temporal approximation is
\[
\begin{aligned}
&\left\{\prod_{t=1}^Tq(\bm\beta_t)q(\bm\eta_t)q(\sigma_t)\right\}
q(\phi_\beta)q(\phi_\eta)q(\sigma_\beta^2)
q(\sigma_{\eta_1}^2)q(\sigma_\eta^2)\\
&\hspace{35mm}\times
\prod_{t=1}^T\prod_{i\in\mathcal S_t}q(\nu_{it}).
\end{aligned}
\]
Each coordinate update has the same distributional family as its Gibbs full
conditional, with quantities involving the other blocks replaced by their
current variational expectations \citep{blei_2017}.

Let $\bar\mu_{it}=E_q(\mu_{it})$,
$\bar r_{it}=y_{it}-\bar\mu_{it}$, and
\[
R_{it}=E_q\{(y_{it}-\mu_{it})^2\}
=\bar r_{it}^2+\operatorname{Var}_q(\mu_{it}).
\]
The variational factor for $\nu_{it}$ is
\[
q(\nu_{it})=\operatorname{GIG}
\left(\frac12,\widetilde\chi_{it},\widetilde\psi_{it}\right),
\]
where the parameters are
\[
\widetilde\chi_{it}=\frac{w_{it}}{\gamma^2}
E_q(\sigma_t^{-1})R_{it}
\qquad\text{and}\qquad
\widetilde\psi_{it}=w_{it}E_q(\sigma_t^{-1})
\left(\frac{\theta^2}{\gamma^2}+2\right).
\]
The known moments of the generalized inverse Gaussian distribution
\citep{Jorgensen1982} give
\[
\begin{aligned}
E_q(\nu_{it})&=\kappa_{it}(1+\omega_{it}^{-1}), &
E_q(\nu_{it}^{-1})&=\kappa_{it}^{-1},\\
E_q(\nu_{it}^2)&=\kappa_{it}^2
(1+3\omega_{it}^{-1}+3\omega_{it}^{-2}).
\end{aligned}
\]
where $\omega_{it}=(\widetilde\chi_{it}\widetilde\psi_{it})^{1/2}$ and
$\kappa_{it}=(\widetilde\chi_{it}/\widetilde\psi_{it})^{1/2}$.

The variational factor for $\sigma_t$ is inverse-gamma with shape
$\widetilde a_{\sigma,t}=a_\sigma+3n_t/2$ and scale
\[
\begin{aligned}
\widetilde b_{\sigma,t}
&=b_\sigma+\sum_{i\in\mathcal S_t}w_{it}E_q(\nu_{it})\\
&\quad+\frac{1}{2\gamma^2}\sum_{i\in\mathcal S_t}w_{it}
\left\{E_q(\nu_{it}^{-1})R_{it}-2\theta\bar r_{it}
+\theta^2E_q(\nu_{it})\right\}.
\end{aligned}
\]
Thus $E_q(\sigma_t^{-1})=\widetilde a_{\sigma,t}/
\widetilde b_{\sigma,t}$.

For the cross-sectional model, define the joint coefficient vector
$\bm\xi=(\bm\beta',\bm\eta')'$ and
\[
\begin{aligned}
\overline D
&=\diag\left\{\frac{w_i}{\gamma^2}
E_q(\sigma^{-1})E_q(\nu_i^{-1})\right\},\\
\bm z_i
&=\frac{w_i}{\gamma^2}E_q(\sigma^{-1})
\{E_q(\nu_i^{-1})y_i-\theta\}.
\end{aligned}
\]
Then $q(\bm\xi)=N(\overline{\bm\xi},\Sigma_\xi)$, where
\begin{align*}
\Sigma_\xi
&=\left(C'\overline DC+\operatorname{blockdiag}
\{\sigma_\beta^{-2}I_q,E_q(\sigma_\eta^{-2})I_r\}\right)^{-1},\\
\overline{\bm\xi}&=\Sigma_\xi C'\bm z.
\end{align*}
The variational factor for $\sigma_\eta^2$ is inverse-gamma with shape
$a_\eta+r/2$ and scale
\[
b_\eta+\frac12
\left\{\overline{\bm\eta}'\overline{\bm\eta}
+\operatorname{tr}(\Sigma_\eta)\right\}.
\]

The Gaussian variational factors for $\bm\beta_t$ and $\bm\eta_t$ use the
same forms as the Gibbs updates above, with quantities involving other
blocks replaced by their variational expectations. In particular,
$q(\bm h_t)=N(\overline{\bm h}_t,\Sigma_{h,t})$, where
\begin{align*}
\Sigma_{h,t}
&=\left(A_t'\overline D_tA_t+
E_q(\lambda_{h,t})I_{d_h}\right)^{-1},\\
\overline{\bm h}_t
&=\Sigma_{h,t}\{A_t'\overline{\bm z}_{h,t}
+E_q(\bm b_{h,t})\}.
\end{align*}
Here $\overline D_t$ is defined as above within year $t$. The vector
$\overline{\bm z}_{h,t}$ uses $E_q(\bm\eta_t)$ when
$\bm h_t=\bm\beta_t$ and $E_q(\bm\beta_t)$ when
$\bm h_t=\bm\eta_t$. For example, when $\bm h_t=\bm\beta_t$, its $i$th element is
\[
\frac{w_{it}}{\gamma^2}E_q(\sigma_t^{-1})
\left[E_q(\nu_{it}^{-1})
\{y_{it}-\bm\psi_{it}'\overline{\bm\eta}_t\}-\theta\right].
\]
The corresponding expression for $\bm\eta_t$ replaces the spatial term by
$\bm x_{it}'\overline{\bm\beta}_t$.

For either $\bm h_t=\bm\beta_t$ or $\bm h_t=\bm\eta_t$, define
\[
\begin{aligned}
\overline S_h
&=\sum_{t=2}^T E_q(\bm h_{t-1}'\bm h_{t-1}), &
\overline R_h
&=\sum_{t=2}^T
\overline{\bm h}_t'\overline{\bm h}_{t-1}.
\end{aligned}
\]
The variational factor for $\phi_h$ is
\[
q(\phi_h)=N_{(-1,1)}
\left(\frac{\overline R_h}{\overline S_h},
\{E_q(v_h^{-1})\overline S_h\}^{-1}\right).
\]
Its first two moments are computed from the standard moments of a truncated
normal distribution. The variational factor for $v_h$ has shape
$a_h+d_h(T-1)/2$ and scale equal to
\[
b_h+\frac12\sum_{t=2}^T
E_q\{\|\bm h_t-\phi_h\bm h_{t-1}\|^2\}.
\]
The expectation in this expression is
\[
\begin{aligned}
E_q\{\|\bm h_t-\phi_h\bm h_{t-1}\|^2\}
&=E_q(\bm h_t'\bm h_t)
-2E_q(\phi_h)\overline{\bm h}_t'\overline{\bm h}_{t-1}\\
&\quad+E_q(\phi_h^2)E_q(\bm h_{t-1}'\bm h_{t-1}).
\end{aligned}
\]
The variational factor for $\sigma_{\eta_1}^2$ has shape
$a_\eta+r/2$ and scale
\[
b_\eta+\frac12E_q(\bm\eta_1'\bm\eta_1).
\]

\subsection{Evidence lower bound and convergence}

The coordinate ascent algorithms evaluate the full evidence lower bound
after every sweep. This calculation is part of the fitted algorithm and is
not an optional diagnostic. For either model,
\[
\mathcal L(q)=E_q\{\log p(\bm y,\bm\nu,\bm\vartheta)\}
-E_q\{\log q(\bm\nu,\bm\vartheta)\},
\]
where $\bm\vartheta$ contains the model parameters.
For each observation, define
\begingroup\small
\[
\begin{aligned}
\ell_{it}
={}&-\frac12\log(2\pi\gamma^2)
+\frac32\log w_{it}-\frac32E_q\{\log(\sigma_t)\}\\
&-w_{it}E_q(\sigma_t^{-1})
\left\{E_q(\nu_{it})+
\frac{E_q(\nu_{it}^{-1})R_{it}-2\theta\bar r_{it}
+\theta^2E_q(\nu_{it})}{2\gamma^2}\right\}.
\end{aligned}
\]
Also let
\[
\begin{aligned}
\mathcal H_{it}^{\nu}
={}&\log 2+\log K_{1/2}(\omega_{it})
+\frac14\log\left(\frac{\widetilde\chi_{it}}
{\widetilde\psi_{it}}\right)\\
&+\frac12\left\{
\widetilde\chi_{it}E_q(\nu_{it}^{-1})
+\widetilde\psi_{it}E_q(\nu_{it})\right\}.
\end{aligned}
\]
\endgroup
The contribution of the asymmetric Laplace mixture representation is then
\[
\mathcal L_{\mathrm{AL}}
=\sum_{t=1}^T\sum_{i\in\mathcal S_t}
\left(\ell_{it}+\mathcal H_{it}^{\nu}\right).
\]
The term $\mathcal H_{it}^{\nu}$ is the entropy of $q(\nu_{it})$.

For clarity, write the remaining terms using inverse-gamma,
Gaussian, and truncated-normal contributions. If the variational factor is
$q(Z)=\operatorname{IG}(\widetilde a,\widetilde b)$ and the prior is
$p(Z)=\operatorname{IG}(a,b)$, then
\begin{align*}
E_q\{\log p(Z)\}
&=a\log b-\log\Gamma(a)
-(a+1)E_q\{\log Z\}-bE_q(Z^{-1}),\\
H\{q(Z)\}
&=\widetilde a+\log\widetilde b+
\log\Gamma(\widetilde a)
-(1+\widetilde a)\psi(\widetilde a).
\end{align*}
For $q(\bm z)=N(\bm m,V)$ of dimension $d$,
\[
H\{q(\bm z)\}=\frac12\{d(1+\log 2\pi)+\log|V|\}.
\]
\par\noindent
If $q(\phi)=N_{(-1,1)}(m,v)$, let
$a=(-1-m)/\sqrt v$, $b=(1-m)/\sqrt v$, and
$Z_\phi=\Phi(b)-\Phi(a)$. Its entropy is
\[
H\{q(\phi)\}=\frac12\log(2\pi e v)+\log Z_\phi
+\frac{a\varphi(a)-b\varphi(b)}{2Z_\phi},
\]
where $\varphi$ and $\Phi$ are the standard normal density and distribution
functions.

For the cross-sectional model, the complete ELBO is
\[
\begin{aligned}
\mathcal L_{\mathrm{CS}}
={}&\mathcal L_{\mathrm{AL}}
+E_q\{\log p(\bm\beta)\}
+E_q\{\log p(\bm\eta\mid\sigma_\eta^2)\}
+E_q\{\log p(\sigma_\eta^2)\}
+E_q\{\log p(\sigma)\}\\
&+H\{q(\bm\xi)\}
+H\{q(\sigma_\eta^2)\}
+H\{q(\sigma)\}.
\end{aligned}
\]
\par\noindent
For the temporal model, it is
\begingroup\small
\[
\begin{aligned}
\mathcal L_{\mathrm{T}}
={}&\mathcal L_{\mathrm{AL}}
+E_q\{\log p(\bm\beta_1)\}
+\sum_{t=2}^TE_q\{\log p(\bm\beta_t\mid\bm\beta_{t-1},
\phi_\beta,\sigma_\beta^2)\}\\
&+E_q\{\log p(\bm\eta_1\mid\sigma_{\eta_1}^2)\}
+\sum_{t=2}^TE_q\{\log p(\bm\eta_t\mid\bm\eta_{t-1},
\phi_\eta,\sigma_\eta^2)\}\\
&+E_q\{\log p(\phi_\beta)\}
+E_q\{\log p(\phi_\eta)\}
+E_q\{\log p(\sigma_\beta^2)\}
+E_q\{\log p(\sigma_{\eta_1}^2)\}
+E_q\{\log p(\sigma_\eta^2)\}\\
&+\sum_{t=1}^TE_q\{\log p(\sigma_t)\}
+\sum_{t=1}^T\left[H\{q(\bm\beta_t)\}
+H\{q(\bm\eta_t)\}+H\{q(\sigma_t)\}\right]\\
&+H\{q(\phi_\beta)\}+H\{q(\phi_\eta)\}
+H\{q(\sigma_\beta^2)\}
+H\{q(\sigma_{\eta_1}^2)\}
+H\{q(\sigma_\eta^2)\}.
\end{aligned}
\]
\endgroup
\par\noindent
The Gaussian prior expectations in these two expressions are evaluated from
the same first and second moments used in the coordinate updates. The
uniform priors for $\phi_\beta$ and $\phi_\eta$ each contribute $-\log 2$.

We monitor the absolute change in ELBO between successive sweeps. Iterations
stop when
\[
|\mathcal L^{(k)}-\mathcal L^{(k-1)}|<10^{-4}.
\]

\subsection{MCMC convergence diagnostics for domain estimates}
\label{sec:mcmc_prediction_diagnostics}

We used four separately initialized chains, each run for 8,000 iterations with the first 1,000 discarded as burn-in.
For posterior prediction, we retained every seventh post-burn-in iteration, giving 1,000 draws from each chain.
The diagnostics cover the state $\times$ year estimates reported at $p=0.50$ and the state $\times$ age $\times$ race $\times$ year estimates reported at all three quantile levels.

We calculated the rank-normalized split $\widehat R$, bulk effective sample size (ESS), and tail ESS proposed by \citet{vehtari_etal_2021} using the \texttt{posterior} package \citep{burkner_etal_2026}.
Following \citet{vehtari_etal_2021}, values of $\widehat R\leq 1.01$ and ESS $\geq 400$ provide useful screening thresholds.
Because the analysis produces thousands of domain estimates, Table~\ref{tab:prediction_diagnostics} reports the maximum $\widehat R$ and the percentages of estimates with $\widehat R>1.01$ or bulk or tail ESS below 400.

\begin{table}[ht!]
\caption{MCMC convergence diagnostics for the finite-population domain estimates reported in the data analysis, based on four chains and 1,000 retained draws per chain.}
\label{tab:prediction_diagnostics}
\centering
\begingroup
\small
\setlength{\tabcolsep}{3pt}
\renewcommand{\arraystretch}{1.15}
\begin{tabular}{llrrrr}
\toprule
$p$ & Aggregation (num. cells) & \shortstack{Max $\widehat R$} & \shortstack{$\widehat R$\\$>1.01$} & \shortstack{Bulk ESS\\$<400$} & \shortstack{Tail ESS\\$<400$} \\
\midrule
0.10 & State $\times$ age $\times$ race $\times$ year ($13{,}824$) & 1.025 & 1.34\% & 0.56\% & 0\% \\
0.50 & State $\times$ year ($384$) & 1.012 & 0.78\% & 0\% & 0\% \\
     & State $\times$ age $\times$ race $\times$ year ($13{,}824$) & 1.011 & 0.04\% & 0\% & 0\% \\
0.90 & State $\times$ age $\times$ race $\times$ year ($13{,}824$) & 1.011 & 0.05\% & 0.01\% & 0\% \\
\bottomrule
\end{tabular}
\endgroup
\end{table}

At $p=0.10$, the maximum $\widehat R$ is 1.025, but only 185 of the 13,824 estimates, or 1.34\%, exceed 1.01.
The maximum $\widehat R$ is no greater than 1.012 for the other reported estimates, with 3 of 384 state $\times$ year estimates at $p=0.50$ and no more than 0.05\% of the remaining cell-level estimates exceeding 1.01.
Bulk ESS is below 400 for 78 of the 13,824 estimates at $p=0.10$ and one estimate at $p=0.90$.
All tail ESS values exceed 400.

\end{document}